# Observation of nuclear-spin Seebeck effect

T. Kikkawa[1,2,3,*], D. Reitz[4], H. Ito[1], T. Makiuchi[1], T. Sugimoto[1], K. Tsunekawa[1], S. Daimon[1], K. Oyanagi[3,5], R. Ramos[2,‡], S. Takahashi[2], Y. Shiomi[6], Y. Tserkovnyak[4], and E. Saitoh[1,2,3,7,8]

1. Department of Applied Physics, The University of Tokyo, Tokyo 113-8656, Japan.

2. WPI Advanced Institute for Materials Research, Tohoku University, Sendai 980-8577, Japan.

3. Institute for Materials Research, Tohoku University, Sendai 980-8577, Japan.

4. Department of Physics and Astronomy, University of California, Los Angeles, California 90095, USA.

5. Faculty of Science and Engineering, Iwate University, Morioka 020-8551, Japan.

6. Department of Basic Science, The University of Tokyo, Tokyo 153-8902, Japan.

7. Advanced Science Research Center, Japan Atomic Energy Agency, Tokai 319-1195, Japan.

8. Institute for AI and Beyond, The University of Tokyo, Tokyo 113-8656, Japan.

‡ Present address: *Centro de Investigación en Química Biolóxica e Materiais Moleculares (CIQUS), Departamento de Química-Física, Universidade de Santiago de Compostela, Santiago de Compostela 15782, Spain.*

*Correspondence and requests for materials should be addressed to T.K. (email: t.kikkawa@ap.t.u-tokyo.ac.jp).

**Abstract**

**Thermoelectric effects have been applied to power generators and temperature sensors that convert waste heat into electricity. The effects, however, have been limited to electrons to occur, and inevitably disappear at low temperatures due to electronic entropy quenching. Here, we report thermoelectric generation caused by nuclear spins in a solid: nuclear-spin Seebeck effect. The sample is a magnetically ordered material $MnCO_3$ having a large nuclear spin ($I = 5/2$) of $^{55}Mn$ nuclei and strong hyperfine coupling, with a Pt contact. In the system, we observe low-temperature thermoelectric signals down to 100 mK due to nuclear-spin excitation. Our theoretical calculation in which interfacial Korringa process is taken into consideration quantitatively reproduces the results. The nuclear thermoelectric effect demonstrated here offers a way for exploring thermoelectric science and technologies at ultralow temperatures.**

**Introduction**

Thermoelectric effects enable the direct conversion of thermal energy into electric energy, promising for power generation and waste heat recovery. Most of the prevalent thermoelectric generators have relied on the Seebeck effect, which is the generation of an electric voltage by placing a conductor junction in a temperature gradient[1-3]. Recently, in the study of spintronics, a spin analogue of the Seebeck effect – the spin Seebeck effect (SSE)[4-20] – was discovered. The SSE is the generation of a spin current, a flow of spin angular momentum, as a result of a temperature gradient applied across a junction consisting of a magnet and a metal[17]. In electronic SSE, a thermally-generated magnon flow in a magnet injects a conduction-electron spin current into the adjacent metal via the interfacial electronic spin exchange[8,9,13,16]. The spin current injected into a metal can be converted into a voltage by the inverse spin Hall effect (ISHE)[21-24], enabling unexplored approaches toward thermoelectric conversion and energy harvesting technologies[10,17,18].

Up to now, all the thermoelectric effects have been an exclusive feature of electrons[1-20]. At low temperatures, however, their efficiency is dramatically suppressed since the thermodynamic entropy of electrons steeply reduces to zero when approaching absolute zero temperature, according to the third law of thermodynamics. In the case of Seebeck effects in semiconductors, the entropy reduction is related to the exponential suppression of the thermally-excited charge carriers[2], while, in SSEs, it is related to the freezing-out of spin fluctuations (magnons)[15-17]. Seebeck effects in metals are also suppressed at low temperatures since the efficiency is governed by $k_\text{B}T/\epsilon_\text{F}$ (ref.[1]), where $k_\text{B}$ is the Boltzmann constant, $T$ the environmental temperature, and $\epsilon_\text{F}$ the Fermi energy. Therefore, so far, the thermoelectric applications have been limited to higher temperatures since no mechanism in the ultralow temperature regime (~ mK range) has been found.

In solids, there is a hitherto unexplored entropy carrier that is well activated even at ultralow temperatures: a nuclear spin. Because of its tiny gyromagnetic ratio $\gamma_\text{n}$ (~ $10^3$ times less than that of electrons[25] $\gamma_\text{e}$), a nuclear spin exhibits much lower excitation energy than that of electron spins in ambient fields, allowing its thermal agitation. Here, a question arises: Can nuclear spins generate thermoelectric effects? If spin angular momentum can be extracted from nuclei in the form of an electron spin current under a temperature bias, it should generate a thermoelectric voltage via the ISHE in an attached metal, realizing all-solid-state thermoelectricity based on atomic nuclei.

Here, we report observation of the nuclear-spin Seebeck effect (Fig. 1a) in a heterostructure composed of a Pt film and a crystal of easy-plane canted antiferromagnetic $MnCO_3$ (refs.[26-28]; Fig. 1d). In $MnCO_3$, $^{55}Mn$ nuclei, a 100% natural-abundance isotope, carry a large spin $I$ of 5/2 and exhibit strong hyperfine coupling with electrons, which allows spin transfer between nuclei and electrons as recently found in the spin pumping measurements under nuclear magnetic resonance[28]. In $MnCO_3$ single crystals covered with Pt films, we found a strong thermoelectric signal enhancement down to 100 mK (Fig. 1e), as shown below, which demonstrates thermoelectric generation at ultralow temperatures. The experimental results are quantitatively reproduced by a theory for nuclear SSE in which the Korringa process[29] due to the hyperfine coupling between nuclear spins in the $MnCO_3$ and conduction electron spins in the attached Pt is taken into consideration (Fig. 1a).

**Results**

**Sample and measurement setup**

We have used the inverse spin Hall effect (ISHE)[21-24] in the Pt film to detect a spin current injected into the film. The ISHE converts a spin current, $\mathbf{J}_\text{s}$, into an electric field, $\mathbf{E}_\text{ISHE}$, through the spin-orbit interaction of conduction electrons, which can be strong in heavy metals such as Pt (refs.[10,17]). When a spin current induced by a nuclear SSE carries spin polarization $\hat{s}$ parallel to the net nuclear-spin polarization **I** along the spatial direction **J**$_s$, **E**$_{ISHE}$ is given by (Fig. 1b)

$$\mathbf{E}_{ISHE} = \frac{2e}{\hbar}\rho\theta_{SHE}\,\mathbf{J}_s\times\hat{s}, \tag{1}$$

where $\rho$ and $\theta_{SHE}$ are the resistivity and the spin Hall angle of the Pt layer, respectively. By measuring **E**$_{ISHE}$, nuclear SSEs can be detected electrically. We note that, since the spin current **J**$_s$ flows normal to the Pt/MnCO$_3$ interface (**J**$_s$ ∥ **x**), the resultant voltage signal $V$ is maximal for $\hat{s}$ ( ∥ **I**) ∥ **z**, when **E**$_{ISHE}$ is measured along the **y** direction shown in Fig. 1b. However, because of the tiny Zeeman coupling of nuclear spins, it is challenging to control nuclear-spin polarization by using **B**, unlike electronic magnetization in conventional magnets. Nevertheless, we can overcome the difficulty by using a magnetic ordered material carrying a large nuclear spin and strong hyperfine coupling. We have noticed that an antiferromagnet MnCO$_3$ ($I$ = 5/2)[26-28] satisfies all such conditions. Below the Néel temperature ($T_N$ = 35 K) of MnCO$_3$, the Mn$^{2+}$ sublattice magnetizations **M**$_1$ and **M**$_2$ are aligned in the (111) plane and canted slightly from the collinear antiferromagnetic configuration due to the bulk Dzyaloshinskii–Moriya interaction[26] (see Fig. 1d and Supplementary Note 1). The hyperfine (Overhauser) fields **B**$_{hf}$ acting on the $^{55}$Mn sublattice nuclear spins **I**$_1$ and **I**$_2$ due to **M**$_1$ and **M**$_2$ reach as large as 57 T (ref.[28]), which induce nuclear spin polarization (~ 40% at 100 mK) and orient **I**$_1$ and **I**$_2$ along the **M**$_1$ and **M**$_2$ directions, respectively[30], as shown in Fig. 1d. Moreover, the net nuclear-spin polarization (**I**$_1$ and **I**$_2$) direction can be controlled by applying **B** since the canting angle $\theta$ of **M**$_1$ and **M**$_2$ changes with **B** owing to the very weak magnetocrystalline anisotropy (~ 0.1 mT within the easy plane[26], see Fig. 1d). The advantage enables us to prepare a controllable nuclear spin polarization in MnCO$_3$, making nuclear SSE experiments feasible.

The SSE devices used in the present study consist of a 10-nm-thick Pt strip [200 μm long ($l$) and 100 nm wide ($w$)] deposited on the top of an insulating MnCO$_3$ (111) (3×3×0.5 mm$^3$) crystal (see Methods and Supplementary Note 2). The Pt strip acts as a heater as well as a spin-voltage converter based on the ISHE for measuring nuclear SSEs: by applying an a.c. current $I_c$ (=$\sqrt{2}I_{rms}\sin\omega t$) to the Pt strip to generate heat and measuring the second harmonic voltage $V$ generated in the Pt by a lock-in technique[11,14], we can selectively detect the ISHE voltage arising from the temperature drop across the Pt/MnCO$_3$ interface induced by the Joule heating $\propto I^2_{\text{rms}}$ of the applied current. The SSE experiments were conducted with a $^4$He cryostat down to 1.82 K using a Pt/MnCO$_3$ device named Device 1 and with a $^3$He–$^4$He dilution refrigerator down to 100 mK using a similar device named Device 2. The Pt/MnCO$_3$ devices were mounted in the cryostats and the magnetic field **B** was applied along the **z** direction as shown in Fig. 1b. Further details are described in Methods.

**Observation of nuclear-spin Seebeck effect**

In Fig. 2a, we show the voltage $V$ data measured at $T = 20$ K and 1.82 K for the Pt/MnCO$_3$ Device 1. At 20 K, no voltage signal appears with the application of $B$. On the other hand, at a lower temperature $T = 1.82$ K, an unconventional voltage signal shows up. The sign of $V/I^2_{\text{rms}}$ reverses by reversing the $B$ direction. The signal intensity increases monotonically with increasing $B$ from zero, and it takes a broad peak at around 4 T. For further high $B$, $V/I^2_{\text{rms}}$ starts to decrease. We confirmed that the observed signal shares the characteristic feature of ISHE induced by SSE[10,17]; $V$ appears only when a heat current is applied and the $V$ intensity scales linearly with the heat power $\propto I^2_{\text{rms}}$. The signal intensity is maximal when **B** ∥ **z**, but vanishes when **B** ⊥ **z**, consistent with the prediction of Eq. (1). The sign of $V$ reverses when the Pt strip ($\theta_{\text{SHE}} > 0$) is replaced with tungsten exhibiting a negative[10] $\theta_{\text{SHE}}$. The results confirm that the voltage signal is induced by thermally driven spin currents and ISHE (see Supplementary Notes 3 to 5 for details).

Surprisingly, the signal intensity persists down to the ultralow temperature regime. Figures 2c and 2d show the $B$ dependence of $V/I^2_{\text{rms}}$ at $1.8\,\text{K} < T < 50\,\text{K}$ for Device 1 and at $100\,\text{mK} < T < 1.6\,\text{K}$ for Device 2, respectively. With decreasing temperature $T$ starting from 50 K, the SSE signal appears below ~10 K and its intensity dramatically increases by further decreasing $T$ (see Figs. 2c and 2b, in which the $T$ dependence of the maximum $V/I^2_{\text{rms}}$ is plotted). Importantly, the signal intensity continues to increase down to ultralow temperatures on the order of ~ 100 mK (see Fig. 2d and the inset to Fig. 2b). Moreover, the signal persists in the higher field range up to 14 T even at such ultralow temperatures, which is totally distinct from the conventional SSE driven by electronic magnetization dynamics. For instance, in ferrimagnetic $Y_3Fe_5O_{12}$, the SSE intensity decreases monotonically with decreasing temperature below 20 K and completely disappears below 5 K at 14 T due to the freezing-out of magnons[15-17]. The maximum output of Device 2 normalized by its electrical resistance $R_{Pt}$, heating power $R_{Pt}I_{rms}^2$, and geometric factor $l^{-1}$ is as large as $V_{max}l/(R_{Pt}^2I_{rms}^2) \sim 58$ nAmW$^{-1}$ at 101 mK, which is nearly two orders of magnitude higher than that of a prototypical room-temperature SSE device made of Pt/$Y_3Fe_5O_{12}$ (~ 1 nAmW$^{-1}$) having the same electrode and heater dimensions [see Supplementary Note 7 and Eq. (3) in Supplementary Note 9 for details].

**Nuclear- and electron-spin excitation spectra in MnCO$_3$**

We now discuss the results in terms of the nuclear- and electron-spin excitation features in MnCO$_3$. In Fig. 1c, we show the electronic and nuclear spin-excitation spectra in MnCO$_3$ (refs.[26-28]) for several fields at $T = 100$ mK, whose thermal energy $k_BT$ is depicted as the green dashed line. Above $k_BT$, thermal excitation is exponentially suppressed. The lower branch at around 600 MHz, corresponding to 30 mK, originates from the nuclear spin excitation $\omega_n$, whose excitation gap is dominated by the strong hyperfine internal field $B_{hf} = \omega_n/\gamma_n \sim 57$ T. The upper branches represent the electronic spin-wave modes $\omega_{mk}$, which shift toward higher frequencies with increasing $B$ due to the strong Zeeman effect. At $B = 14$ T, the electronic spin excitation gap $\omega_{m0}(\approx \gamma_e B)$ is ~ 19 K, two orders of magnitude greater than the thermal energy = 100 mK, resulting in a negligibly small value of the Boltzmann factor $\exp(-\hbar\omega_{m0}/k_BT) \sim 10^{-82} \ll 1$. If the SSE we measured were driven by the electronic spin-wave modes, the SSE signal would be completely suppressed by applying a strong field of 14 T, as with the conventional SSE of $Y_3Fe_5O_{12}$ (ref.[15]). This clearly shows the irrelevance of the electronic SSE to the observed signal at low temperatures. On the other hand, the nuclear spin mode can be greatly excited even by such a small thermal energy of ~ 100 mK, and it remains almost unaffected by the applied $B$ due to the tiny Zeeman effects, much weaker than the hyperfine internal field ~ 57 T (Fig. 1c); the nuclear spins can contribute to SSEs even in such a low-$T$ and high-$B$ environment. The results also suggest that direct coupling between nuclear spins in the MnCO$_3$ and electrons in the Pt at the interface should be responsible for the SSE, rather than the interfacial electronic exchange mediated by the gapped magnons under strong magnetic fields.

**Theoretical model for nuclear-spin Seebeck effect**

We theoretically model the nuclear SSE in which direct nuclear-electron coupling due to the Korringa process[29] is taken into consideration. In the model, the spin current $J_{ne}$ is generated by the interfacial hyperfine interaction between nuclear spins in the MnCO$_3$ and conduction electron spins in the Pt under the temperature bias $T_e - T_p$ (see Fig. 3a and Supplementary Note 9 for details). Here, $T_e$ and $T_p$ represent effective temperatures for electrons in the Pt and phonons in the MnCO$_3$ near the interface, respectively. The nuclear spin current $J_{ne}$ arises in proportion to the effective temperature difference between the electrons in the Pt ($T_e$) and nuclei in the MnCO$_3$ ($T_n$): $J_{ne} = \Gamma_{ne} k_B (T_e - T_n)$. Here, $T_n$ may deviate from the electron $T_e$ due to the nuclear-phonon thermalization in MnCO$_3$ given by $J_{np} = \Gamma_{np} k_B (T_n - T_p)$, resulting in the finite spin current $J_{ne}$. The expression for the nuclear SSE coefficient reads

$$S_n = \frac{g_n^{\uparrow\downarrow}}{4\pi I} \pi \chi b \frac{\hbar \omega_n}{k_B T} \left[ \frac{T_e - T_n}{T_e - T_p} \right] \qquad (2)$$

where $g_n^{\uparrow\downarrow}$ is the nuclear spin-mixing conductance per unit area, $\chi$ the normalized antiferromagnetic transverse susceptibility such that $\theta = \chi b$ is the canting angle, $b \equiv \hbar \gamma_e s_e B$ the normalized magnetic field with saturated spin density $s_e$ ($s_e \equiv S/V$, for volume per site $V$), and $T$ the average temperature. The bracketed expression in Eq. (2) is evaluated as $(T_e - T_n)/(T_e - T_p) = (1 + \Gamma_{ne}/\Gamma_{np})^{-1}$ from the steady-state condition $J_{ne} = J_{np}$ (ref.[31]). Here, $\Gamma_{ne} \propto 1/T$ [Eq. (1) shown in Supplementary Note 9] and $\Gamma_{np} \propto 1/T\omega_{m0}^2$ is derived by Fermi's Golden rule for the nuclear-phonon thermalization rate mediated by virtual magnons (see Supplementary Note 9), which allow us to evaluate the $B$ dependence of $T_e - T_n$. As shown in Fig. 3b, it is maximal at zero field by the strong thermalization (i.e., $T_n \sim T_p$) and decreases gradually with $B$. There is a crossover field $B_c$, marked by $\Gamma_{np}$ falling below $\Gamma_{ne}$ (see the results at $T = 100$ mK and 1 K in Fig. 3b). In Fig. 3c, we compare the $B$ dependence of the experimental $V/I_{rms}^2$ (blue plots) for Device 2 and calculated $V/I_{rms}^2$ based on the nuclear SSE $S_n$ (red solid curve) at $T = 100$ mK. Of important note, the experimental data is quantitatively reproduced by the calculation. Such agreement is confirmed also for other $B$ and $T$ regions (see Figs. 3d and 3e). A non-monotonic field response of $V$ now becomes evident: for $B \ll B_c$, the SSE signal increases in proportion to $B$ ($S_n \propto B$) owing to the increased canting angle, and it takes a maximum at $B \sim B_c$. For $B \gg B_c$, the SSE signal decreases monotonically with $B$ ($\mathcal{S}_\text{n} \propto B^{-1}$) due to the reduction of thermal nonequilibrium $T_\text{e} - T_\text{n}$ ($\propto B^{-2}$) between the electron and nuclear systems (see Fig. 3b). We also evaluated the electronic SSE, $\mathcal{S}_\text{m}$, driven by the antiferromagnetic spin-wave mode $\omega_{\text{m}k}$ (see Supplementary Note 9 for details) and found that its intensity as well as $B$ and $T$ dependencies do not explain the experimental results (see Fig. 3c and its inset), which confirm that the nuclear SSE dominates the observed SSE.

**Discussion**

We finally discuss the difference between the previous nuclear spin pumping[28] and the present nuclear SSE. For the nuclear spin pumping, the measured voltage is maximal at a relatively low field of ~ 0.3 T and then starts to decrease with $B$. In such a low-$B$ range, the excitation gap of electronic spin-wave branch in MnCO$_3$ is comparable to that of the nuclear spins, and a nuclear-spin wave, hybridized electronic spin-wave and nuclear spin mode[32-34], is excited. The experimental result in ref.[28] was thereby attributed to the coherent nuclear spin-wave formation, the electronic (magnetization) component of which pumps a spin current into an adjacent metallic layer in analogy with the conventional electronic spin pumping for a magnet/metal bilayer. On the other hand, the present nuclear SSE increases with $B$ up to around 4 – 5 T, while nuclear-electronic hybridization is quickly suppressed as the electronic spin-waves become gapped out. This suggests that a different physical mechanism governs the nuclear SSE, which is reasonable as the nuclear pumping in the SSE is not limited to a coherent long-wavelength dynamics. We thus develop a nuclear SSE theory in terms of interfacial Korringa relaxation, in which nuclear-spin fluctuation directly transmits a spin current into an attached metallic layer via interfacial hyperfine interaction, and found quantitative agreement between the experiment and calculation. The Korringa mechanism does not need strong nuclear-electronic spin hybridization in the magnetic layer and also electronic spin transfer at the interface. This may extend a class of materials applicable for nuclear spintronics; materials having magnetic elements with nuclear spins and strong hyperfine interaction, such as $^{55}$Mn and $^{59}$Co (both of which are 100% natural abundance), can be potential sources of nuclear spin currents.

In summary, we demonstrated the thermoelectric conversion driven by nuclear spin: the nuclear SSE. The nuclear SSE is enhanced at ultralow temperatures, in stark contrast to conventional electron-based thermoelectricity. It is surely worthwhile to explore nuclear SSEs in other systems to show the generality of the phenomenon. Materials of interest include easy-axis antiferromagnetic insulators having a large nuclear spin and exhibiting a spin-flop transition, at which the electronic magnon gap comes close to the low energy scales relevant to the nuclear dynamics[35]; for the nuclear SSE, this is instrumental in thermal equilibration of the nuclei within the magnetic material.

The present work may serve as the bridge between nuclear-spin science and thermoelectricity and marks the beginning of a research field "Nuclear thermoelectricity". It is also worth exploring the reciprocal of the nuclear SSE, since it will be applied to making a nuclear heat pump working at ultralow temperatures.

## Methods

### Sample preparation

We used single-crystalline $MnCO_3$ slabs with a size of $3\times3\times0.5$ mm$^3$, which are commercially available from SurfaceNet. The largest plane is (111) in the rhombohedral representation[36,37]. On the top of the (111) plane of the $MnCO_3$ slabs, 10-nm-thick Pt strips (200 μm long and nominally 100 nm wide) were patterned by electron beam lithography and lift-off methods (see also Supplementary Note 2). The Pt strips were deposited by magnetron sputtering in a $10^{-1}$ Pa Ar atmosphere. For a control experiment, we also prepared W/$MnCO_3$ devices, where the Pt strips are replaced with 10-nm-thick W strips (200 μm long and 500 nm wide) exhibiting a negative $\theta_{SHE}$ (refs.[10,38,39]).

### Spin Seebeck effect measurement

We measured the SSE by a standard lock-in technique[11,14,40,41] with a PPMS (Quantum Design) from 1.8 K to 50 K and a $^3$He–$^4$He dilution refrigerator (KelvinoxMX200, Oxford Instruments; cooling power of 200 μW at 100 mK) from 100 mK to 10 K. An a.c. charge current ($I_c = \sqrt{2}I_{rms}\sin\omega t$) was applied to the Pt strip with a current source (6221, Keithley) and the generated voltage $V$ across the strip was recorded with a lock-in amplifier (LI5640, NF Corporation). For the measurements with the dilution refrigerator, we further introduced a voltage preamplifier (1201, DL Instruments) and a programmable filter (3625, NF Corporation) to reduce signal noise. The typical a.c. charge current property: the root-mean-square (rms) amplitude $I_{rms}$ of 0.1 - 5 μA, and the frequency $\omega/2\pi$ of 13.423 Hz. All the $V$-$B$ data are anti-symmetrized with respect to the magnetic field $B$.

**Data availability**

The data that support the findings of this study are available from the corresponding author upon reasonable request.

**Code availability**

The codes used in theoretical simulations and calculations are available from the corresponding authors upon reasonable request.

**Acknowledgements**

This is a post-peer-review, pre-copyedit version of an article published in Nature Communications. The final authenticated version is available online at: http://dx.doi.org/10.1038/s41467-021-24623-6.We thank Y. Chen, J. Lustikova, T. Hioki, N. Yokoi, H. Chudo, M. Imai, K. Sato, and G. E. W. Bauer for fruitful discussions and T. Nojima for his valuable comments on low-temperature experiments. This work was supported by JST ERATO "Spin Quantum Rectification Project" (JPMJER1402), JST CREST (JPMJCR20C1 and JPMJCR20T2), JSPS KAKENHI (JP19H05600, JP19K21031, JP20H02599, JP20K22476, and JP20K15160), MEXT (Innovative Area "Nano Spin Conversion Science" (JP26103005)), and Daikin Industries, Ltd. The work at UCLA was supported by the US Department of Energy, Office of Basic Energy Sciences under Award No. DE-SC0012190. K.O. acknowledges support from GP-Spin at Tohoku University. R.R. acknowledges support from the European Commission through the project 734187-SPICOLOST (H2020-MSCA-RISE-2016), the European Union's Horizon 2020 research and innovation program through the Marie Sklodowska-Curie Actions grant agreement SPEC number 894006 and the Spanish Ministry of Science (RYC 2019-026915-I).

**Author contribution**

T.K., T.S., and K.O. fabricated the devices. T.K. and T.M. constructed the experimental setup with the help of R.R. and K.O. T.K., T.M., H.I., T.S., K.T., and S.D. performed the experiments and collected the data. T.K. and H.I. analyzed the data with input from D.R. D.R. and Y.T. developed the theoretical explanations. E.S. and Y.T. conceived and supervised the project. T.K. and D.R. wrote the paper with review and input from E.S. and Y.T. T.K., D.R., H.I., T.M., T.S., K.T., S.D., K.O., R.R., S.T., Y.S., Y.T., and E.S. discussed the results and commented on the manuscript.

**<u>Competing interests</u>**

The authors declare no competing interests.

**<u>Correspondence</u>** Correspondence and requests for materials should be addressed to T.K. (email: t.kikkawa@ap.t.u-tokyo.ac.jp).

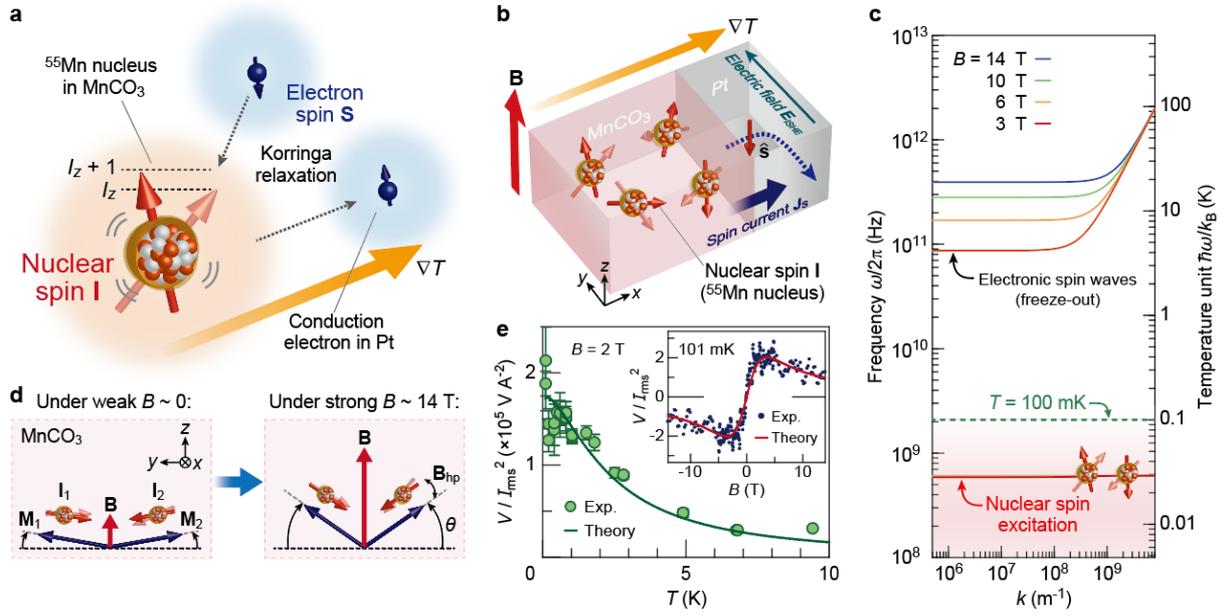

**Fig. 1 | Concept of nuclear-spin Seebeck effect in Pt/MnCO$_3$. a**, Schematic illustration of the nuclear SSE induced by the Korringa relaxation process[29], the spin-conserving flip-flop scattering between a nuclear spin, **I**, of $^{55}$Mn in MnCO$_3$ and an electron spin, **S**, in Pt via the interfacial hyperfine interaction. $I_z$ represents the $z$ component of the nuclear spin **I**. **b**, Schematic illustration of the nuclear SSE and the ISHE in a Pt/MnCO$_3$ hybrid structure, where MnCO$_3$ contains nuclear spin $I$ = 5/2 on $^{55}$Mn (100% natural abundance). When a temperature gradient ($\nabla T$) is applied across the Pt/MnCO$_3$ interface, a spin current (**J**$_s$) carrying a spin polarization vector $\hat{\mathbf{s}}$ is induced in the Pt layer by the nuclear SSE, which is subsequently converted into an electric field (**E**$_{ISHE}$) via the ISHE in the direction of $\mathbf{J}_s \times \hat{\mathbf{s}}$ (ref.[22]). Here, $\hat{\mathbf{s}}$ is along the external magnetic field **B**. **c**, A calculated dispersion relation of the electronic spin wave and energy of the nuclear spin excitation in MnCO$_3$ at a temperature $T$ of 100 mK for several magnetic fields[26-28]. The energy level of 100 mK is plotted with a green dashed line. At $T$ = 100 mK, the electronic spin waves are frozen out, while nuclear spins remain thermally active. **d**, Schematic illustration of the orientation of the Mn$^{2+}$ sublattice electronic magnetization **M**$_1$ and **M**$_2$ and the $^{55}$Mn nuclear spins **I**$_1$ and **I**$_2$ in MnCO$_3$ in the (111) plane when the external field **B** is applied in the plane (**B** ∥ **z**). Below the antiferromagnetic ordering temperature $T_N$ = 35 K of MnCO$_3$, **M**$_1$ and **M**$_2$ are aligned in the (111) plane and canted slightly from the pure antiferromagnetic ordering direction due to the bulk Dzyaloshinskii–Moriya interaction[26] (Supplementary Fig. 1). The canting angle $\theta$ increases with the external field. $\theta$ = 0.26° at zero field, while $\theta$ = 12° at $B$ = 14 T. Due to the strong hyperfine (Overhauser) field of $B_{hf}$ ~ 57 T, the sublattice nuclear spins **I**$_1$ and **I**$_2$ orient antiparallel to the electronic magnetization **M**$_1$ and **M**$_2$ directions, respectively. Here, the antiparallel orientation originates from the nature of the contact hyperfine interaction and the relative sign of the nuclear and

electronic gyromagnetic ratios $\gamma_n$ and $\gamma_e$ (ref.[30]). **e**, Experimental demonstration of the nuclear SSE in Pt/MnCO$_3$. Temperature ($T$) dependence of the thermoelectric voltage $V$ (normalized by the applied heat power $\propto I_{rms}^2$) at $B$ = 2 T. The error bar represents the standard deviation. The inset shows the $B$ dependence of $V/I_{rms}^2$ at $T$ = 101 mK. Theoretical results for the nuclear SSE are also plotted with solid curves.

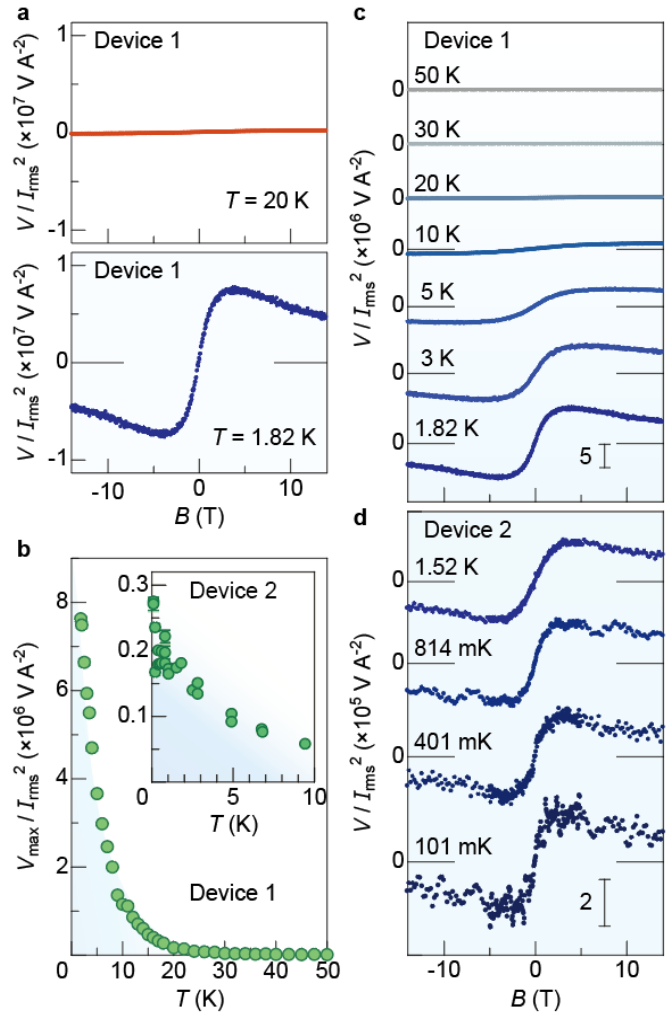

**Fig. 2 | Observation of nuclear-spin Seebeck effect in Pt/MnCO₃. a**, $B$ dependence of $V/I_{rms}^2$ (voltage $V$ normalized by the square of the applied charge current $I_{rms}$) for the Pt/MnCO₃ Device 1 at $T = 20$ K (red) and 1.82 K (blue). **b**, $T$ dependence of the maximum $V/I_{rms}^2$ (defined as $V_{max}/I_{rms}^2$) for the Pt/MnCO₃ Device 1 at 1.8 K < $T$ < 50 K. The inset shows the $T$ dependence of $V_{max}/I_{rms}^2$ for the Pt/MnCO₃ Device 2 at 100 mK < $T$ < 10 K measured with a dilution refrigerator. The error bar represents the standard deviation. **c, d**, $B$ dependence of $V/I_{rms}^2$ for the Pt/MnCO₃ Devices 1 (**c**) and 2 (**d**) at selected temperatures. The Pt/MnCO₃ Device 1 exhibits electrical resistance one order of magnitude higher than that for Device 2, resulting in an overall higher intensity of $V/I_{rms}^2$ in the Pt/MnCO₃ Device 1 (see Supplementary Note 6 for details).

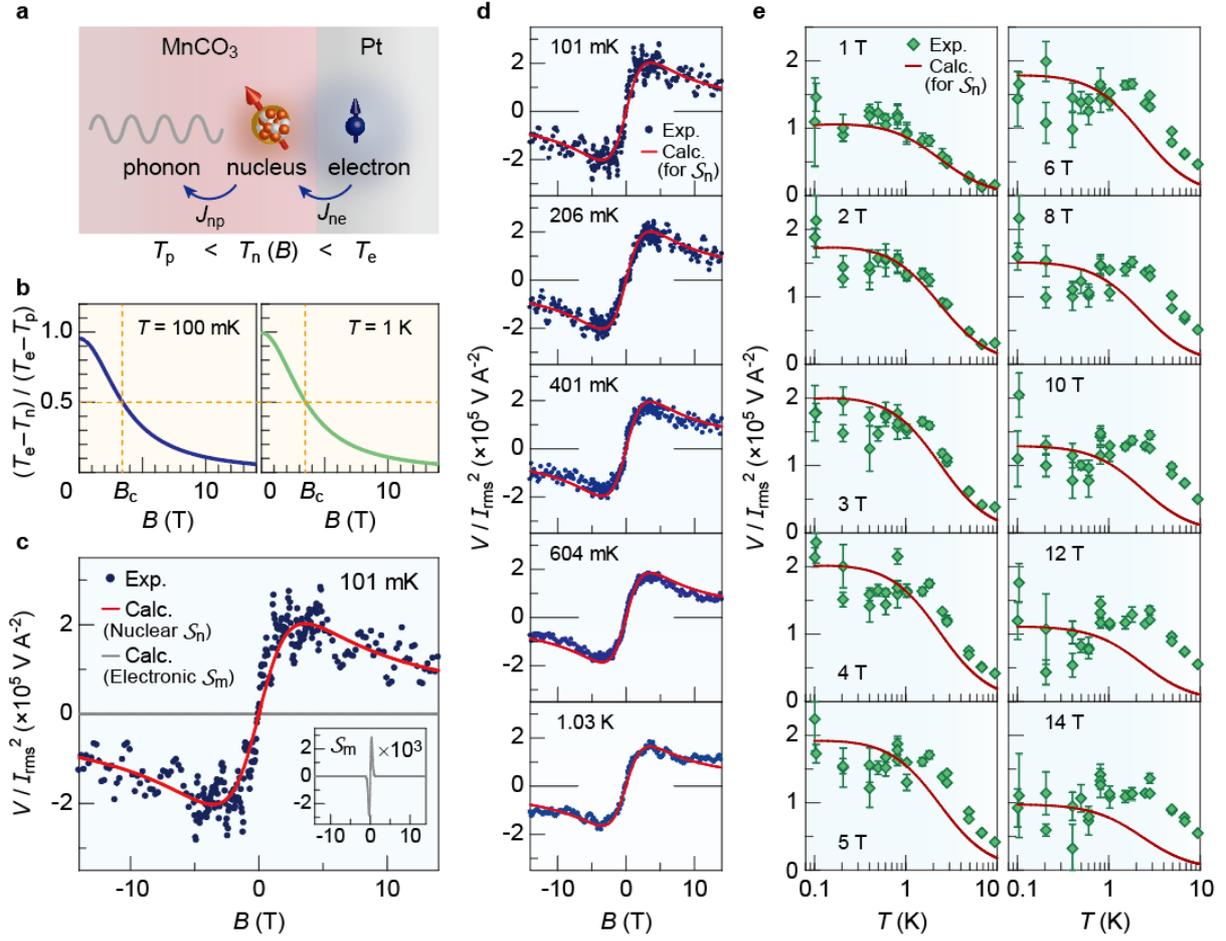

**Fig. 3 | Comparison between experiment and theory. a**, Interfacial nuclear spin current and thermal equilibration of nuclear spins in MnCO$_3$. An interfacial spin current, $J_{ne}$, is mediated by the Korringa process through the hyperfine interaction between nuclear spins of $^{55}$Mn and electron spins in the metal at the Pt/MnCO$_3$ interface. $J_{ne}$ arises in proportion to the effective temperature difference between the electrons in Pt ($T_e$) and nuclei in MnCO$_3$ ($T_n$): $J_{ne} = \Gamma_{ne} k_B (T_e - T_n)$. Here, the difference $T_e - T_n$ may be triggered by the interfacial temperature drop $T_e - T_p$ between the Pt and MnCO$_3$ ($T_p$: phonon temperature in MnCO$_3$ close to the interface) and the thermalization between nuclei and phonons in MnCO$_3$, whose rate is given by $J_{np} = \Gamma_{np} k_B (T_n - T_p)$. **b**, $B$ dependence of the calculated temperature difference $T_e - T_n$ normalized by the interfacial temperature drop $T_e - T_p$ at $T$ = 100 mK and 1 K. In the steady state, $J_{ne} = J_{np}$ (ref.[31]), which gives $(T_e - T_n)/(T_e - T_p) = \Gamma_{np}/(\Gamma_{np} + \Gamma_{ne})$. $B_c$ indicates the crossover field, where $\Gamma_{ne} = \Gamma_{np}$. **c**, Comparison between the $B$ dependence of the experimental $V/I_{rms}^2$ (blue plots) for the Pt/MnCO$_3$ Device 2 and the calculated $V/I_{rms}^2$ for the nuclear SSE $S_n$ (red solid curve) and for the electronic SSE $S_m$ (gray solid curve) at $T$ = 101 mK (see Supplementary Note 9 for details). The inset shows a blowup of the calculated $V/I_{rms}^2$ for the electronic SSE $S_m$ (multiplied by $10^3$). **d**, Comparison

between the *B* dependence of the experimental $V/I_{rms}^2$ (blue plots) and the calculated $V/I_{rms}^2$ for the nuclear SSE $S_n$ (red solid line) at 100 mK < *T* < 1 K. **e**, Comparison between the *T* dependence of the experimental $V/I_{rms}^2$ (green rhombus) and the calculated $V/I_{rms}^2$ for the nuclear SSE $S_n$ (red solid curve). The error bar represents the standard deviation.

# Supplementary Information for
# Observation of nuclear-spin Seebeck effect

T. Kikkawa et al.,

**Table of Contents:**

**Supplementary Note 1 | Magnetization measurement**

The magnetization $M$ of the MnCO$_3$ slab was measured using the Vibrating Sample Magnetometer (VSM) option of a Physical Properties Measurement System (PPMS, Quantum Design) in the temperature range from 2 K to 300 K. An external magnetic field, **B**, was applied parallel to the (111) plane of the MnCO$_3$. The $M$-$T$ curve of the MnCO$_3$ at $B = 25$ mT for $T <$ 60 K is shown in Supplementary Fig. 1, from which the antiferromagnetic ordering (Néel) temperature $T_N$ was estimated to be ~ 35 K.

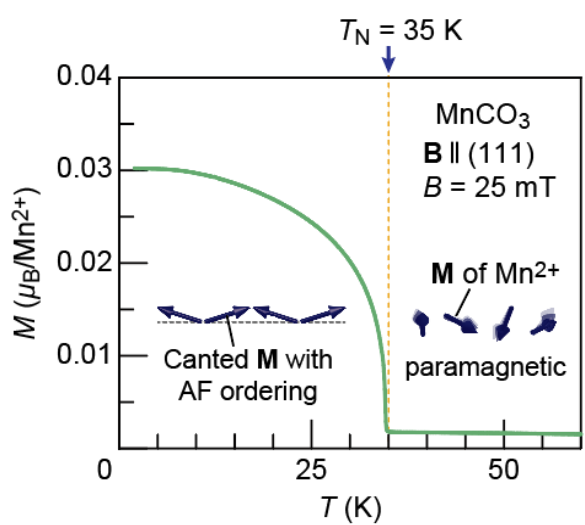

**Supplementary Fig. 1 | Magnetic property of MnCO$_3$.** Temperature ($T$) dependence of the magnetization $M$ ($M$-$T$ curve) of the MnCO$_3$ slab, measured by applying the external magnetic field $B = 25$ mT parallel to the (111) plane of MnCO$_3$. With decreasing $T$, the $M$ abruptly increases at around $T_N = 35$ K. This is due to the antiferromagnetic (AF) ordering of Mn$^{2+}$ magnetization **M**, which are canted slightly from the collinear antiparallel alignment because of the bulk Dzyaloshinskii–Moriya interaction, causing a small net magnetization (~ 0.03 $\mu_B$/Mn$^{2+}$) (ref.[1]). The canting angle $\theta$ is around 0.26° at zero field, which can be modulated by changing $B$. At $B = 14$ T, $\theta$ ~ 12°.

**Supplementary Note 2 | Device characterization**

A laser microscope image of a Pt/MnCO$_3$ device, an atomic force microscope image of the (111) surface of the MnCO$_3$, and an X-ray diffraction result of the MnCO$_3$ (111) basal plane are shown in Supplementary Figs. 2a, 2b, and 2c, respectively.

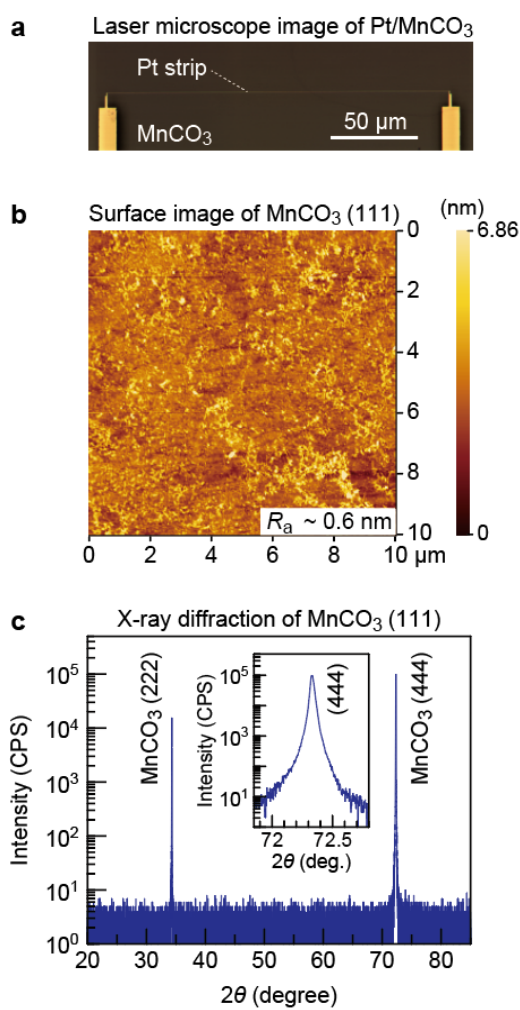

**Supplementary Fig. 2 | Microscope image of Pt/MnCO$_3$ device and surface and crystallinity analyses. a**, A laser microscope image of a Pt/MnCO$_3$ device. A 10-nm-thick Pt strip (200 µm long and 100 nm wide) is patterned by the electron beam lithography. The white scale bar shows 50 µm. **b**, An atomic force microscope image of the polished (111) surface of MnCO$_3$. The averaged roughness $R_a$ is ~ 0.6 nm, showing nice flatness. **c**, A 2$\theta$-$\omega$ X-ray diffraction pattern for a MnCO$_3$ (111) basal plane. CPS in the vertical axis denotes the count per second. Two clear peaks are observed at around 34.2° and 72.3°, which are assigned as the diffraction from the (222) and (444) planes of MnCO$_3$, respectively (in the rhombohedral setting). The inset shows a blowup of the (444) diffraction peak.

## Supplementary Note 3 | Signal characteristics

In Supplementary Figs. 3a and 3b, we show the voltage $V$ signal as a function of the applied heat current ($\propto I_{rms}^2$) in the Pt/MnCO$_3$ Device 1. $V$ appears only when a heat current is applied and the $V$ intensity scales linearly with the heat power $\propto I_{rms}^2$. As shown in Supplementary Fig. 3c, when **B** is rotated in the $xz$-plane at an angle $\alpha$ to the $x$ direction, the $V$ signal varies with $\alpha$ in a $\sin\alpha$ pattern and vanishes when $\alpha = 0°$ and $180°$ (**B** ∥ ±**x**), consistent with the signal characteristic of the inverse spin-Hall effect (ISHE) induced by the spin Seebeck effect (SSE). We also found that the sign of $V$ reverses when the Pt strip ($\theta_{SHE} > 0$) is replaced with tungsten exhibiting a negative spin Hall angle $\theta_{SHE}$ (refs.[2-4]; see Supplementary Fig. 4).

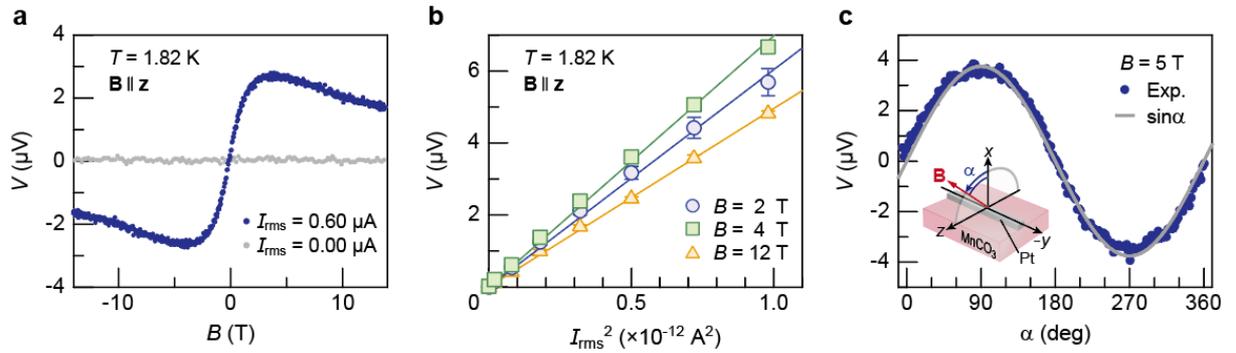

**Supplementary Fig. 3 | Voltage characteristics. a**, $B$ dependence of $V$ for the Pt/MnCO$_3$ Device 1 at $T$ = 1.82 K for the applied current intensity of $I_{rms}$ = 0.60 µA (blue plots) and 0 (gray plots), where the magnetic field **B** is applied along the **z** direction ($\alpha$ = 90°). The $V$ signal disappears in the absence of the applied current. **b**, $I_{rms}^2$ dependence of $V$ for the same device at $T$ = 1.82 K for several $B$ values (**B** ∥ **z**). The error bar represents the standard deviation. **c**, The magnetic-field angle $\alpha$ dependence of $V$ for the same device at $T$ = 1.82 K and $I_{rms}$ = 0.71 µA (blue plots), where the external field of $B$ = 5 T is rotated in the $x$-$z$ plane. $\alpha$ denotes the angle between the $x$ axis and the **B** direction. The gray solid curve is a $\sin\alpha$ fit. For the $\alpha$ dependence measurement, the Pt/MnCO$_3$ sample is rotated under a fixed magnetic field $B$ with use of a horizontal rotator option of PPMS.

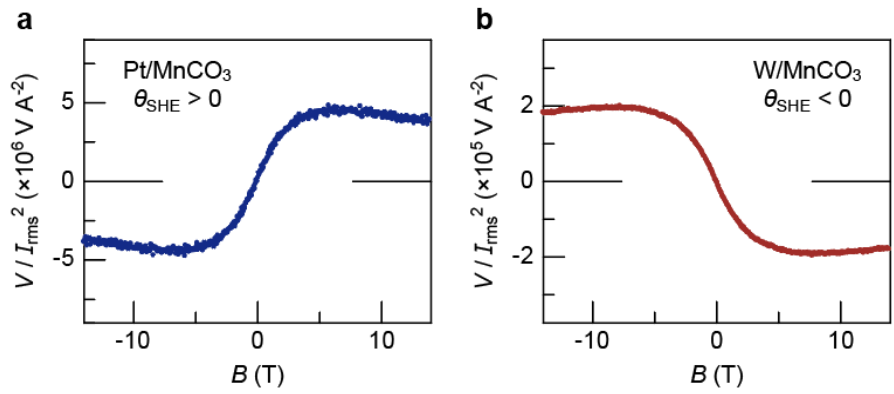

**Supplementary Fig. 4 | Comparison of the voltage signal between Pt/MnCO$_3$ and W/MnCO$_3$. a,b**, $B$ dependence of $V/I_{rms}^2$ at $T$ = 4 K for the Pt/MnCO$_3$ Device 1 (**a**, blue plots) and for the W/MnCO$_3$ device (**b**, red plots). The observed $V$ signal changes its sign when Pt is replaced with W. $\theta_{SHE}$ > 0 for Pt and $\theta_{SHE}$ < 0 for W (refs.[2-4]).

**Supplementary Note 4 | Separation between nuclear-spin Seebeck effect and uni-directional spin Hall magnetoresistance**

In our experimental setup for nuclear SSE measurements, an a.c. charge current is applied directly to the Pt wire and the resultant second harmonic voltage $V$ is detected as a function of $B$. This setup is almost identical to that for the unidirectional spin Hall magnetoresistance (USMR) caused by a current-induced spin accumulation[5-11], whose voltage signal also scales with the applied current squared $I_{rms}^2$ and shows a $B$-asymmetric dependence as with the SSE. This means that the separation between the nuclear SSE and USMR is of importance. To separate them, we have performed control experiments and found that the nuclear SSE indeed dominates the observed signal, as elaborated below.

To this end, we first conducted SSE measurements using a chip-heater/Pt/MnCO$_3$ structure, where a resistive chip heater is attached on the Pt layer. In this setup, by applying a charge current to the chip heater (not the Pt layer), a temperature gradient can be created across the Pt/MnCO$_3$ interface, allowing examination of the nuclear SSE free from the possible USMR contribution. Supplementary Fig. 5a shows the $B$ dependence of the second harmonic voltage $V$ generated in the Pt layer at $T = 2$ K measured with a lock-in technique. Voltage signals show up clearly, whose sign changes with respect to the $B$ reversal. Besides, as shown in Supplementary Fig. 5b, the $V$ intensity scales with the applied heating power $R_h I_{rms}^2$, where $R_h$ and $I_{rms}$ are the resistance of the heater ($R_h = 100$ Ω) and the amplitude of the applied a.c. charge current ($I_c = \sqrt{2} I_{rms} \sin\omega t$), respectively. These features are consistent with the characteristics of the ISHE voltage induced by SSE. Furthermore, we found that the temperature $T$ dependence of the signal in this experimental configuration also agrees with that measured for the Pt/MnCO$_3$ devices shown in Fig. 2b in the main text; the SSE signal appears below ~ 10 K and its intensity monotonically increases by decreasing $T$ (see Supplementary Figs. 5c and 5d). The results show that the nuclear SSE manifests in the absence of the external charge current to the Pt layer, which is free from the possible USMR signal.

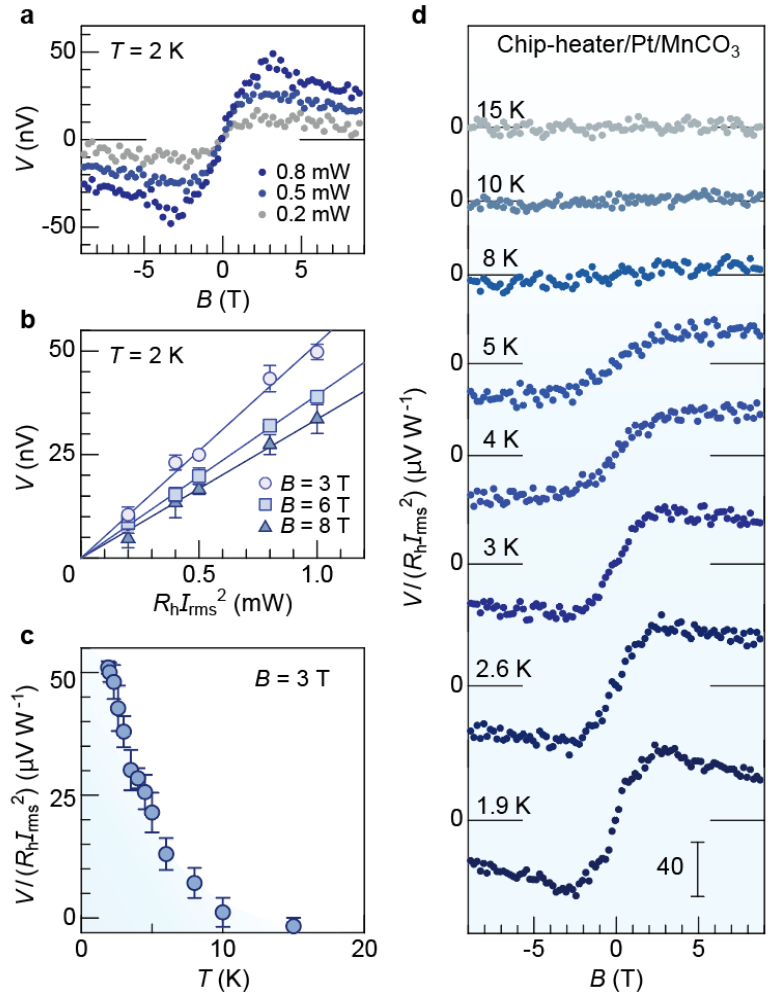

**Supplementary Fig. 5 | Nuclear-spin Seebeck effect in chip-heater/Pt/MnCO$_3$ structure. a**, $B$ dependence of $V$ for the chip-heater/Pt/MnCO$_3$ sample at $T$ = 2 K for several heating power $R_h I_{rms}^2$, where $R_h$ and $I_{rms}$ are the resistance of the heater ($R_h$=100 Ω) and the amplitude of the a.c. charge current, respectively. The sample consists of a 5-nm-thick Pt film sputtered on the whole surface (3×3 mm$^2$) of a MnCO$_3$ (111) slab with a size of 3×3×0.5 mm$^3$. Here, on the top of the Pt layer, a resistive heater is attached to create a temperature gradient across the Pt/MnCO$_3$ interface. **b**, Heating power $R_h I_{rms}^2$ dependence of $V$ for several $B$ values, showing a linear relationship. The error bar represents the standard deviation. **c**, $T$ dependence of $V/(R_h I_{rms}^2)$ (voltage $V$ normalized by the heating power $R_h I_{rms}^2$) at $B$ = 3 T for 1.9 K < $T$ < 15 K. The error bar represents the standard deviation. **d**, $B$ dependence of $V/(R_h I_{rms}^2)$ at selected temperatures.

We then performed another control experiment to evaluate the possible USMR contribution under the experimental setup, where an a.c. charge current ($I_c$) is applied directly to the Pt layer.

To this end, we followed the experimental method introduced by Avci et al.[5,6] and measured the longitudinal ($V_L$) and transverse ($V_T$) voltages of a Hall-bar-shaped Pt/MnCO$_3$ bilayer sample, where $V_L$ and $V_T$ are the second harmonic voltages along the Hall-bar's length ($L$ = 60 μm ∥ **y**) and width ($W$ = 10 μm ∥ **z**) directions, respectively. In the configuration, if present, the USMR may appear along the Hall-bar's length direction when the in-plane magnetic field $B$ is applied perpendicular to the charge current $I_c$ (**B** ∥ **z**), while it disappears along the Hall-bar's width direction for **B** ∥ **y**. On the other hand, in both the configuration, the SSE-induced second harmonic voltages can show up and their signal-intensity ratio $V_L/V_T$ scales with the geometric factor $L/W$ (refs.[5,6]). By experiments, we indeed observed $B$-asymmetric second harmonic voltage signals for both the longitudinal $V_L$ for **B** ∥ **z** and transverse $V_T$ for **B** ∥ **y**, whose intensity is proportional to $I_{rms}^2$. The experimental $V_L/V_T$ ratio was found to be ~ 6.45, which agrees with the geometric factor $L/W$ within ~ 10% accuracy. Therefore, we conclude that most of the second harmonic signal is of SSE origin and not related to the USMR. This conclusion is consistent with the previous similar experiments for a magnetic-insulator Y$_3$Fe$_5$O$_{12}$ (YIG) and Pt bilayer reported by Avci et al.[6].

**Supplementary Note 5 | Comments on possible anomalous Nernst effect induced by magnetic proximity effect due to static hyperfine interaction**

We argue that the anomalous Nernst effect (ANE)[12,13] induced by a magnetic proximity effect (due to static hyperfine interaction) does not explain the observed voltage characteristics. First of all, we observed the non-monotonic $B$ dependence of the voltage, whose intensity increases by increasing $B$ from 0, saturates at around 2 T, and then starts to decrease for further high fields (Fig. 3 in the main text). This non-monotonic behavior is not expected from the ANE scenario as it should monotonically increase with $B$ (ref.[14]) due to the increased nuclear spin polarization along $B$. On the other hand, the observed $B$ response can be well reproduced by our Korringa spin-current scenario combined with the thermalization model (Fig. 3). Besides, the polarity of $V$ changes in response to the sign of the spin Hall angle $\theta_{SHE}$ of metallic layer (Supplementary Fig. 4), consistent with the spin-current scenario, where non-equilibrium spin polarization is created via transverse components of interfacial hyperfine interaction. We thus conclude that the nuclear SSE governs the observed signal, rather than the possible proximity ANE.

**Supplementary Note 6 | Comparison of voltage normalized by Pt resistance, heating power, and geometric factor between Pt/MnCO$_3$ Devices 1 and 2**

In the main text, we show the voltage $V$ normalized by the applied charge current squared $I_{\text{rms}}^2$. As shown in Fig. 2, the obtained $V/I_{\text{rms}}^2$ value for the Pt/MnCO$_3$ Device 1 is one order of magnitude higher than that for Device 2. We attribute the difference mainly to that of the electrical resistance $R_{\text{Pt}}$ of the Pt layer; the $R_{\text{Pt}}$ value for Device 1 at $T = 1.8$ K is ~ 179 k$\Omega$, several times higher than that for Device 2 at the same temperature ($R_{\text{Pt}}$ ~ 30.9 k$\Omega$). The output SSE signal indeed scales with $R_{\text{Pt}}^2 I_{\text{rms}}^2 l^{-1}$, since the electromotive force induced by the ISHE is proportional to $R_{\text{Pt}}$ (ref.[15]) and the input heating power in our experimental setup increases proportionally to $R_{\text{Pt}} I_{\text{rms}}^2$ ($l$ is the Pt length along the electrode direction and is the same between Devices 1 and 2) [for details see Eq. (3) in Supplementary Note 9]. As shown in Supplementary Fig. 6, the $Vl/(R_{\text{Pt}}^2 I_{\text{rms}}^2)$ intensity is almost identical between the Pt/MnCO$_3$ Devices 1 and 2 [$V_{\text{max}} l/(R_{\text{Pt}}^2 I_{\text{rms}}^2)$ ~ 48 nAmW$^{-1}$ for Device 1 at $T = 1.82$ K and ~ 38 nAmW$^{-1}$ for Device 2 at $T = 1.80$ K]. This shows that the quantity $Vl/(R_{\text{Pt}}^2 I_{\text{rms}}^2)$ is a nice benchmark to compare the SSE performance with different samples, in which the information of the device resistance and geometry is taken into account. We would like to note that the magnetoresisitance (MR) ratio of our Pt film is as small as ~ 0.02% up to the $B$ intensity of 14 T at $T = 2$ K (Supplementary Fig. 7), meaning that the heating power $R_{\text{Pt}} I_{\text{rms}}^2$ does not depend on $B$ for the whole $B$ range in the present study.

The difference of $R_{\text{Pt}}$ may originate from that of the Pt width; although it was designed to be 100 nm, the resultant Pt width was found to be ~ 30 nm and ~ 97 nm for Devices 1 and 2, respectively, from atomic force microscopy measurements. This may be due to the different type of electron-beam resist PMMA (polymethyl methacrylate) used for making Devices 1 and 2; Device 1 was made with 950 PMMA "A4" resist, while Device 2 with 950 PMMA "A2" (KAYAKU Advanced Materials, Inc.). According to the data sheet from the provider (ref.[16]), due to their different anisole content, the A4-type resist layer becomes about three times thicker than the A2-type after a spin-coating process. This may cause possible underdose in the electron-beam lithography process for Device 1, resulting in the narrow width for the Pt pattern of Device 1. Nevertheless, we would like to note that the Pt resistance difference can be taken into account in our analysis and does not affect our conclusion.

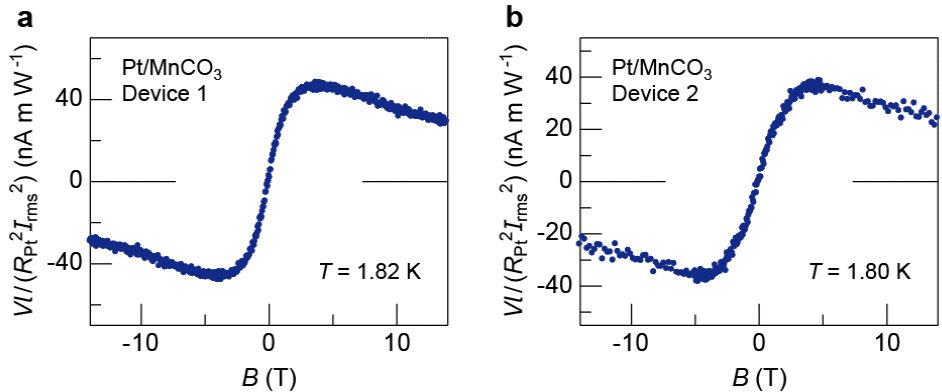

**Supplementary Fig. 6 | Comparison of signal intensity between Pt/MnCO₃ Devices 1 and 2. a,b**, $B$ dependence of $VI/(R_{Pt}^2 I_{rms}^2)$ (voltage $V$ normalized by the Pt resistance $R_{Pt}$, heating power $R_{Pt} I_{rms}^2$ applied to the Pt wire, and the inverse of the Pt length along the electrode direction $l^{-1}$) for the (**a**) Pt/MnCO₃ Device 1 at $T$ = 1.82 K and (**b**) Device 2 at $T$ = 1.80 K.

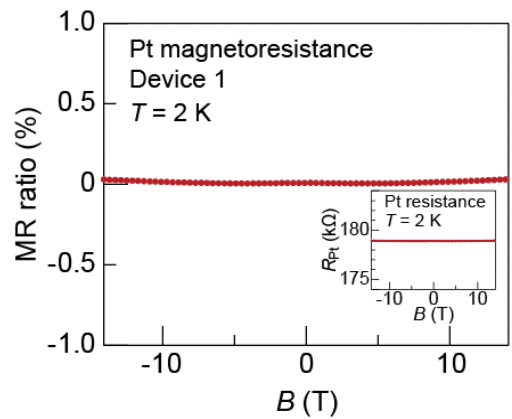

**Supplementary Fig. 7 | Magnetoresistance of Pt wire.** $B$ dependence of the magnetoresistance (MR) ratio relative to zero field for the Pt wire of Device 1 at $T$ = 2 K, where the in-plane $B$ is applied perpendicular to the Pt wire. The inset shows the $B$ dependence of electric resistance $R_{Pt}$ of the Pt wire at the same temperature.

**Supplementary Note 7 | Comparison of voltage normalized by Pt resistance, heating power, and geometric factor between Pt/MnCO$_3$ Device 2 and Pt/Y$_3$Fe$_5$O$_{12}$ (YIG) device**

We compare the $Vl/(R_{Pt}^2 I_{rms}^2)$ value between the Pt/MnCO$_3$ Devices 2 at $T$ = 101 mK and a Pt/YIG-film device at 300 K having the same electrode and heater dimensions. Here, the single-crystalline YIG-film with the thickness of ~ 4 μm is grown on a Gd$_3$Ga$_5$O$_{12}$ (111) substrate with liquid phase epitaxy. The maximum output $V_{max}l/(R_{Pt}^2 I_{rms}^2)$ value is ~ 58 nAmW$^{-1}$ for the Pt/MnCO$_3$ Device 2 as shown in Supplementary Fig. 8a, which is nearly two orders of magnitude higher than that for the Pt/YIG (~ 1 nAmW$^{-1}$; Supplementary Fig. 8b).

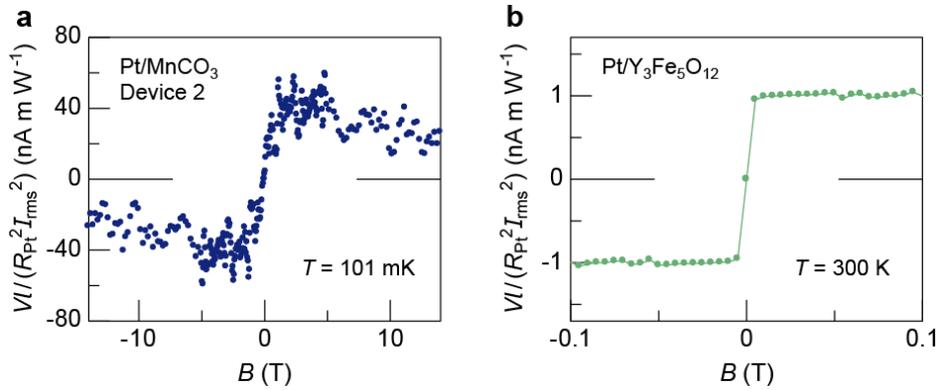

**Supplementary Fig. 8 | Comparison of signal intensity between Pt/MnCO$_3$ Device 2 and Pt/Y$_3$Fe$_5$O$_{12}$. a**, $B$ dependence of $Vl/(R_{Pt}^2 I_{rms}^2)$ for the Pt/MnCO$_3$ Device 2 at $T$ = 101 mK. **b**, $B$ dependence of $Vl/(R_{Pt}^2 I_{rms}^2)$ for the Pt/YIG-film at $T$ = 300 K. Here, the YIG-film with the thickness of ~ 4 μm is grown on a Gd$_3$Ga$_5$O$_{12}$ substrate with liquid phase epitaxy. The Pt layer for the Pt/MnCO$_3$ Device 2 and Pt/YIG has the same dimensions.

**Supplementary Note 8 | Evaluation of temperature difference and nuclear-spin Seebeck thermopower of Pt/MnCO$_3$**

For the nuclear SSE measurement on the Pt/MnCO$_3$ Device 2 at the lowest temperature of $T =$ 101 mK, the intensity of applied a.c. charge current $I_{\text{rms}}$ is as small as 70 nA, which gives the heating power of $R_{\text{Pt}} I_{\text{rms}}^2 = 154$ pW. This value is several orders of magnitude smaller than the cooling power of our dilution fridge: 200 µW at 100 mK. Indeed, with the application of such small heating power during the measurement, there was no change of the sample's temperature monitored by a RuO$_x$ sensor, which was attached to the back of the oxygen-free high thermal conductivity (OFHC grade) copper plate, on which the Pt/MnCO$_3$ sample was mounted. We evaluated the temperature difference $\Delta T$ generated in the sample to be of the order of ~ 0.3 mK by a heat-flow analysis[17], where the Pt wire is modelled as a heat source embedded in the center of a half cylinder of MnCO$_3$ (with a radius of 0.5 mm) exhibiting a thermal conductivity value of $8.8 \times 10^{-6}$ Wcm$^{-1}$K$^{-1}$ at 100 mK which is estimated from the experiment by Ozhogin et al.[18].

Using the above estimated temperature difference $\Delta T \sim 0.3$ mK, we extract the spin-Seebeck thermopower $S$ defined as $S = (V/L_V)/(\Delta T/L_T)$ (ref.[19]), which is most frequently used in the SSE research community. Here, $L_V$ and $L_T$ represent the length along the electrode direction and the thickness of a SSE device, respectively. For the Pt/MnCO$_3$ Device 2 at $T = 101$ mK, $S$ is evaluated as ~ 11 µV K$^{-1}$, which is nearly two orders of magnitude higher than that for 10-nm-thick Pt/YIG-slab junctions measured at room temperature ($S \sim 0.2 – 0.5$ µV K$^{-1}$) (refs.[12,13]), a situation consistent with the comparison shown in Supplementary Fig. 8.

## Supplementary Note 9 | Theoretical model of nuclear-spin Seebeck effect and comparison with experiment

The electronic SSE is driven by both an interfacial temperature difference between magnons and electrons in the metal[20] and a non-equilibrium chemical potential which builds up at the interface due to a bulk thermal gradient[21-24]. While electronic magnons can transport spin due to their finite group velocities and lifetimes, nuclear spin transport is limited to small fields since they have almost zero group velocity aside from the small hybridized region of the Brillouin zone, as can be seen from Fig. 1c in the main text. Furthermore, since here the magnetic field is perpendicular to the Néel order (and the nuclei align locally with the electronic spins), spin is not conserved along the field. Thus, we formulate the nuclear SSE in terms of interfacial thermal spin pumping rather than bulk spin transport.

The measured nuclear SSE voltage, normalized by the input energy flux, reported here appears to be larger than one might expect when compared to, for example, YIG at room temperature in a device with the same geometry. There are several quantities relevant to the scaling of the SSE with temperature: the metal's resistivity, spin-diffusion length, and spin Hall angle; the thermal conductivities, phonon inelastic equilibration and phonon-spin carrier relaxation lengths[25]; and the magnetic dynamics itself via the spin Seebeck coefficient $\mathcal{S}$. At low temperatures the phonons become essentially noninteracting and thus equilibrate very poorly to a common temperature; then for a fixed energy flux, the bulk thermal gradient and temperature drop across the interface increase due to smaller thermal bulk and interfacial conductivities. Additionally, the interfacial temperature drop may be effectively larger due to a longer inelastic phonon equilibration length. Furthermore, since the nuclear spins are all thermalized $\mathcal{S}_n \propto 1/T$ as compared to the magnonic SSE which freezes out as $\mathcal{S}_m \propto T^{3/2}$ for ferromagnets (both bulk and interfacial contributions[26]) and $\mathcal{S}_m \propto T^3$ for this antiferromagnet (see below). Comparing the nuclear SSE contribution to ferromagnetic and antiferromagnetic contributions, at $T \gg T_C$ or $T_N$ we would expect the paramagnetic contributions to show a similar $1/T$ scaling, at which point it should dominate over the nuclear's since interfacial electron exchange coupling is much stronger than the interfacial nuclear hyperfine coupling ($T_C$ and $T_N$ are the Curie and Néel temperatures, respectively). From $T \sim T_C$ or $T_N$ and down, however, the magnonic SSE is reduced, while the nuclear one keeps increasing as $1/T$.

Each nuclear spin aligns antiparallel to the electronic sublattice magnetization on the same site, and the net spin current across two adjacent sites cancels unless there is canting of the electronic spins (with angle $\theta = \chi b$, see Fig. 1d in the main text). We calculate the nuclear spin current transferred into the metal due to Korringa-like relaxation[27] (depicted in Fig. 1a) by Fermi's Golden rule in the limit $k_B T \gg \hbar \omega_n$:

$$J_n = \chi b J_{ne} = \rho(\epsilon_F)^2 a^2 \pi \chi b \hbar \omega_n (T_e - T_n)/T \tag{1}$$

where $J_{ne}$ is the spin current per site and $J_n$ is the average over a pair of sites, $\rho(\epsilon_F)$ is the density of states at the Fermi level in units of (energy·volume)$^{-1}$, $a$ is the interfacial hyperfine interaction constant between nuclei and the spin density in the metal, and $T$ is the average temperature. The temperature dependence in $J_{ne}$ differs from the usual Korringa spin-relaxation rate, $\tau_k^{-1} \propto T$, since $J_{ne}$ is due to the spin flow, minus the backflow, into the Fermi gas. We define the nuclear-electron spin Seebeck coefficient per site $\Gamma_{ne}$ by $J_{ne} \equiv \Gamma_{ne} k_B (T_e - T_n)$ and define the nuclear spin-mixing conductance per unit area as $g_n^{\uparrow\downarrow} \equiv 4\pi s_n \rho(\epsilon_F)^2 a^2$ for saturated nuclear spin density $s_n$ ($s_n \equiv I/\mathcal{A}$, calculated for spin $I = 1/2$ and interfacial area $\mathcal{A}$ per site), in analogy with the electronic result in ref.[28].

The nuclear spin Seebeck coefficient $\mathcal{S}_n$ relates the experimentally-applied thermal bias $T_e - T_p$ to the spin current density $J_n/\mathcal{A} \equiv \mathcal{S}_n k_B (T_e - T_p)$ and involves balancing, for each nuclear site at the interface, the spin flow from phonons in the antiferromagnet to electrons in the metal. When the temperature is well above the magnon gap $\omega_{m0}$, the magnons can directly transfer the small energy $\hbar \omega_n$ to the nuclei by two-magnon (Raman-like absorption and remission) scattering processes in the magnon continuum[29]; rapidly equilibrating the phonon, magnon, and nuclear temperatures. At temperatures comparable to and below the magnon gap, the mechanism driving the nuclear-phonon spin current $J_{np} \equiv \Gamma_{np} k_B (T_n - T_p)$ involves a virtual magnon process which mediates a nuclear spin flip by transmitting energy $\hbar \omega_n$ from phonons. This occurs near the Gamma point since $\omega_n$ is small on the scale of the magnon and phonon dispersions. We diagonalize the magnon-phonon Hamiltonian by performing a Bogoliubov transformation which yields two hybridized branches. The hyperfine interaction in the antiferromagnet now becomes a second-order, direct nuclear coupling to the hybridized field operators. Then to lowest order in magnon hybridization within the low-energy branch, we get by Fermi's Golden rule: $\Gamma_{np} \propto 1/T\omega_{m0}^2$, where $\omega_{m0} = \gamma_e\sqrt{B_{a'}^2 + B^2}$ is the magnon gap for easy-axis anisotropy field[30] $B_{a'}$ and perpendicular applied field, both within the easy plane. The remaining $B,T$-independent coefficient $C$ is taken from experiment: $\Gamma_{np}/\Gamma_{ne} = C/\omega_{m0}^2$. We fit $C$ by aligning the experimental and theoretical crossover fields $B_c$, Thus, in our theory $\mathcal{S}_n$ is rate-limited by thermalization between nuclei and electrons when $B < B_c$, and thermalization between nuclei and phonons when $B > B_c$.

The magnonic spin Seebeck coefficient $\mathcal{S}_m$ is defined in the same way as $\mathcal{S}_n$, relating the spin current density $J_m \equiv \mathcal{S}_m k_B(T_e - T_p)$ to the thermal bias[31]. We calculate $\mathcal{S}_m$ semi-classically using the fluctuation-dissipation theorem, assuming high magnon quality factors as in ref.[32], to get

$$\mathcal{S}_m = \frac{g_m^{\uparrow\downarrow}\hbar\chi b}{2\pi s_e k_B} \int \frac{d^3k}{(2\pi)^3} \omega_{mk}\partial_T n_{BE}(\omega_{mk}) \tag{2}$$

where $g_m^{\uparrow\downarrow}$ is the magnonic interfacial spin-mixing conductance per unit area, $s_e$ is the saturated spin density in the antiferromagnet ($s_e \equiv S/V$, for dimensionless spin $S$ and volume $V$ per site), $n_{BE}(\omega_{mk}) = [\exp(\hbar\omega_{mk}/k_B T) - 1]^{-1}$ is the Bose–Einstein distribution function, and $\omega_{mk} = \sqrt{\gamma_e^2(B_{a'}^2 + B^2) + c^2 k^2}$ where $c$ is the speed of antiferromagnetic spin waves at large wave numbers $k$. $\mathcal{S}_m$ may be evaluated analytically in the limit $k_B T \gg \hbar\omega_{mk}$ which gives $\mathcal{S}_m \propto BT^3$. At low temperatures and large fields, magnon thermal populations at $\omega_{m0}$ are exponentially suppressed[31], causing $\mathcal{S}_m$ to decrease monotonically with increasing field. The transition in field behavior is roughly marked by $\hbar\gamma_e B_c \sim k_B T$. Finally, there is an additional type of contribution to the electronic SSE that is due to the hybridization of nuclei and magnons, which is known as nuclear frequency pulling[33]. However in our system this can be shown to decrease more slowly with increasing field than $\mathcal{S}_m$, but to be negligible compared to $\mathcal{S}_n$.

The voltage measured due to the SSE arises from the inverse spin Hall effect associated with the thermally-induced spin current[34]. When the SSE voltage (normalized by the injector current squared) is driven by an interfacial temperature discontinuity, it is given by[32]

$$\frac{V_{\text{SSE}}}{I_c^2} = \frac{\mathcal{S}_i(B,T)}{\kappa^*(T)} \frac{2ek_B}{\hbar} \frac{\lambda^*}{l} R_d R_h \qquad (3)$$

where $\lambda^* \equiv \lambda_{sd}\theta_{\text{SHE}}$ is the effective spin-diffusion length times the detector's spin Hall angle $\theta_{\text{SHE}} \equiv \hbar J_c/2eJ_s$ for lateral charge density $J_c$ and interfacial spin current density $J_s$, with $J_s = J_n/\mathcal{A}$ for the nuclei and $J_s = J_m$ for the magnons, $l$ is the length of the detector, $R_h$, $R_d$ is the resistance in the metal heater and detector, $\kappa^*$ is the effective phononic interfacial Kapitza conductance (which relates the injected heat flux to $T_e - T_p$ at the interface), and $\mathcal{S}_i$ is the local Seebeck coefficient in units of inverse area ($i = $ n, m). In order to fit the overall common factors (including the device resistance) multiplied by $\mathcal{S}_i$ in $V_{\text{SSE}}$, we use the low-field slope $f(T)$ where both theory and experiment are linear in $B$. The slope goes as $f(T) \propto 1/(a + bT^c)$ where $a = 8.4$, $b = 2.2$, and $c = 1.6$. Since at small fields $\mathcal{S}_m \propto BT^3$ and $\mathcal{S}_n \propto B/T$, for the electronic SSE we would need $\kappa_m^*(T) \propto (a + bT^c)T^3$ and for the nuclear SSE we would need $\kappa_n^*(T) \propto (a + bT^c)/T$ to fit the experimental $f(T)$. The theoretical curves plotted in Fig. 3 for comparison to the data are calculated as $\mathcal{S}_n f(T)/f_n(T)$ and $\mathcal{S}_m f(T)/f_m(T)$, where $f_i(T) \equiv \partial \mathcal{S}_i/\partial B(B = 0)$ are the slopes of the theoretical Seebeck coefficients evaluated at zero field.

While our theory for $\mathcal{S}_n$ reproduces most aspects of the measured signal, it slightly underestimates the data at large relative to small fields (Fig. 3e). This might be explained by additional nuclear-phonon thermalization channels such as direct nuclear-phonon coupling or indirect coupling via the second magnon branch associated with Néel excitations out of the easy plane. Since $\mathcal{S}_n$ is limited by $\Gamma_{np}$ at $B > B_c$, our theory would then give a lower bound on $\mathcal{S}_n$ there, which is consistent with experiment, if the additional channels do not decrease as fast with field as our $\Gamma_{np}$. Additional, inelastic channels for nuclear-electron spin transport may also affect the fit of $f_n(T)$ to $f(T)$, which comes from the data at $B < B_c$ where $\Gamma_{ne}$ limits $\mathcal{S}_n$, and the fit parameter $C$ which controls the position of $B_c$. In our theory, the contributions to $\Gamma_{np}$ and $\Gamma_{ne}$ have the same overall temperature dependence giving a $T$-independent $B_c$, which is consistent with the data over a large range of temperatures.